\documentclass{article}

\usepackage{microtype}
\usepackage{graphicx}
\usepackage{subfigure}
\usepackage{booktabs} 
\usepackage{hyperref}

\usepackage[accepted]{icml2023}

\usepackage{amsmath}
\usepackage{amssymb}
\usepackage{mathtools}
\usepackage{amsthm}

\usepackage[capitalize,noabbrev]{cleveref}

\icmltitlerunning{ABB3}

\begin{document}

\twocolumn[
\icmltitle{ABodyBuilder3: Improved and scalable antibody structure predictions}

\icmlsetsymbol{equal}{*}

\begin{icmlauthorlist}
\icmlauthor{Henry Kenlay}{exs}
\icmlauthor{Frédéric A. Dreyer}{exs}
\icmlauthor{Daniel Cutting}{exs}
\icmlauthor{Daniel Nissley}{exs}
\icmlauthor{Charlotte M. Deane}{exs,oxf}
\end{icmlauthorlist}

\icmlaffiliation{exs}{Exscientia, Oxford Science Park, Oxford, OX4 4GE, UK}
\icmlaffiliation{oxf}{Department of Statistics, University of Oxford, Oxford OX1 3LB, UK}

\icmlcorrespondingauthor{Henry Kenlay}{hkenlay@exscientia.co.uk}

\icmlkeywords{Machine Learning, Structure prediction}

\vskip 0.3in
]



\printAffiliationsAndNotice{} 

\begin{abstract}
Accurate prediction of antibody structure is a central task in the design and development of monoclonal antibodies, notably to understand both their developability and their binding properties.
In this article, we introduce ABodyBuilder3, an improved and scalable antibody structure prediction model based on ImmuneBuilder.
We achieve a new state-of-the-art accuracy in the modelling of CDR loops by leveraging language model embeddings, and show how predicted structures can be further improved through careful relaxation strategies.
Finally, we incorporate a predicted Local Distance Difference Test into the model output to allow for a more accurate estimation of uncertainties.
\end{abstract}
\section{Introduction}
Immunoglobulin proteins play a key role in the active immune system, and have emerged as an important class of therapeutics~\cite{antibodies1}.
They are constructed from two heavy and two light chains, separated into distinct domains.
The tip of each of the two antibody binding arms is defined as the variable region, and contains six complementarity-determining regions (CDRs) across the heavy and light chains which make up most of the antigen-binding site. 
As part of an immune response, B cells undergo clonal expansion, which, coupled with somatic hypermutations and recombinations, leads to an accumulation of mutations in the DNA encoding the CDR loops. 
The remaining domains compose the constant region and are primarily involved in effector function.

Understanding the three-dimensional structure of antibodies is critical to assessing their properties~\cite{chungyoun2023ai} and developability~\cite{tap,raybould23}. 
The framework regions connecting the CDR loops are relatively conserved and thus easily predicted from sequence similarity. Similarly, five of the CDR loops tend to cluster along canonical forms~\cite{pyigclassify,wong19} and are thus relatively straightforward to model.
The third loop of the heavy chain (CDRH3), for which the coding sequence is created during the recombination of the V, D, and J gene segments~\cite{doi:10.1128/microbiolspec.mdna3-0041-2014}, is however more challenging due to its much larger sequence and length diversity.
As the CDRH3 loop often drives antigen recognition, e.g.~\cite{cdrh3_2011,cdrh3}, improving the accuracy with which its structure can be predicted from sequence is a key component to advancing rational antibody design.

Experimental protein structure determination remains a costly and slow process~\cite{struct_challenges}, such that only a small fraction of known antibody sequences have experimentally resolved three-dimensional structural information~\cite{sabdab1,sabdab2}.
One approach to circumvent these experimental limitations is through structure prediction methods, which have had immense success in reaching experimental accuracy on general protein structures~\cite{af2,rosettafold,esmfold}.

Structure models are also a necessary element to accurately predict biophysical properties of proteins and advance the field of rational therapeutic design.
Several dedicated tools have emerged to model specifically the variable region of antibodies.
Among them are IgFold~\cite{igfold}, which is based on a language model, DeepAb~\cite{deepab}, which uses an attention mechanism, ABlooper~\cite{ablooper}, which predicts backbone atom positions using a graph neural network, ABodyBuilder~\cite{abodybuilder}, a homology modelling pipeline, and ABodyBuilder2~\cite{abb2}, which uses a modified version of the AlphaFold-Multimer architecture~\cite{afm}.
More recently, xTrimoPGLM-Ab~\cite{xtrimopglm} has shown promising results on antibody structures by combining a General Language Model framework~\cite{glm} with a modified AlphaFold2 architecture.

\begin{figure*}
    \centering
    \includegraphics[width=1.0\linewidth]{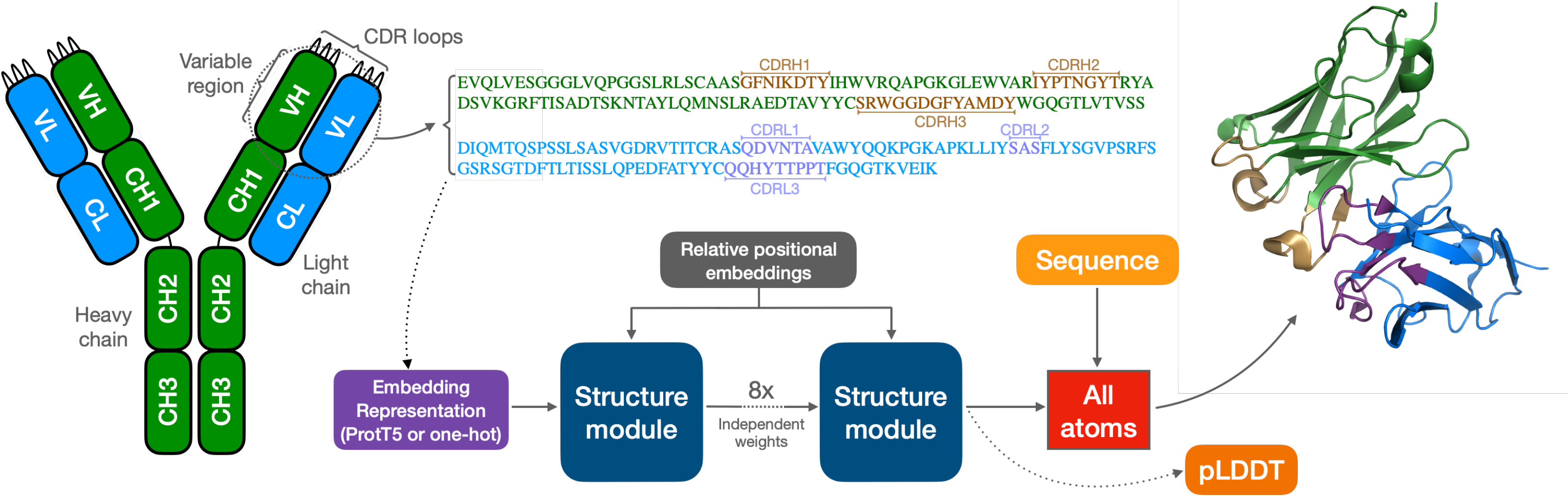}
    \caption{Left: Overview of an antibody structure, with the variable region and CDR loops shown. 
    Right: Schematic representation of the ABodyBuilder3 architecture, with 8 sequential and independent update blocks providing the final atomic coordinates and uncertainty predictions from an embedding representation of the variable region sequence.}
    \label{fig:diagram}
\end{figure*}

In this article, we introduce ABodyBuilder3, an antibody structure prediction model based on ABodyBuilder2~\cite{abb2}.
As shown in Figure~\ref{fig:diagram}, ABodyBuilder2 consists of an embedding representation of the variable region sequence, which is provided as input to a sequence of eight structure modules that update the node features and residue coordinates through an invariant point attention layer and a backbone update layer, respectively.
We detail key changes to the implementation, data curation, sequence representation and structure refinement that improve the scalability and accuracy of the model.
Additionally, we introduce an uncertainty estimation based on the predicted local-distance difference test (pLDDT), which outperforms the previous ensemble-based estimate.
Together, these enhancements provide a substantial improvement in the quality of antibody structure predictions and open the possibility of a scalable and precise assessment of large numbers of therapeutic candidates.

\section{Improved structure modelling and evaluation}

Rapid prototyping is paramount to generating insights and improving the design of machine learning models.
We develop an efficient and scalable implementation of the ABodyBuilder2 architecture which makes use of vectorisation to improve hardware utilisation, leveraging optimisations from the OpenFold project~\cite{ahdritz2022openfold}. This is in contrast to the implementation of ABodyBuilder2, which generates minibatch gradients by computing a gradient for each minibatch sample sequentially before averaging (i.e. accumulated gradients). ABodyBuilder2 also uses double precision which is not well optimised on modern GPU hardware compared to lower precision data-types. We find the model can be trained robustly using bfloat16 precision for weights and use mixed precision for training, providing faster computational throughput and an efficient memory footprint. 
Our implementation is more than three times faster, and can be scaled easily across multiple GPUs using a distributed data parallel strategy. 

\begin{figure*}
    \centering
    \includegraphics[width=0.7\textwidth,trim={1.5cm 1cm 2cm 1.5cm},clip]{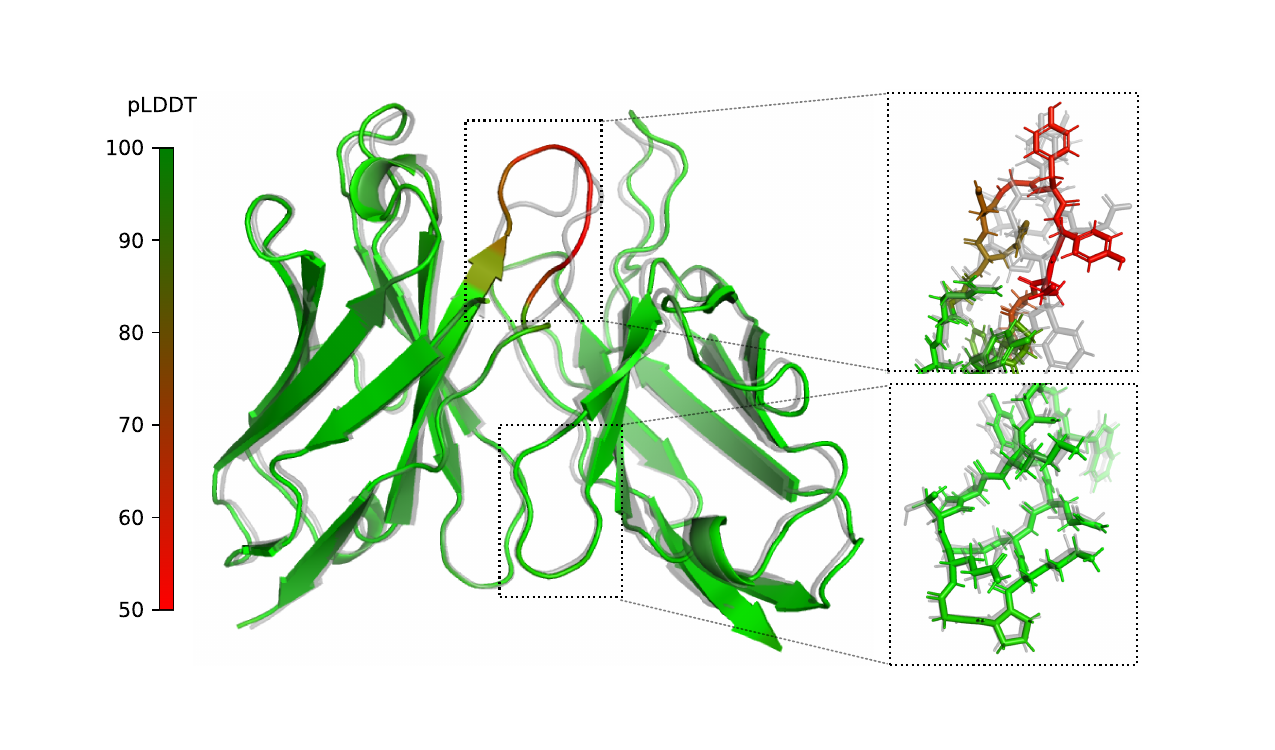}%
    \includegraphics[width=0.28\linewidth]{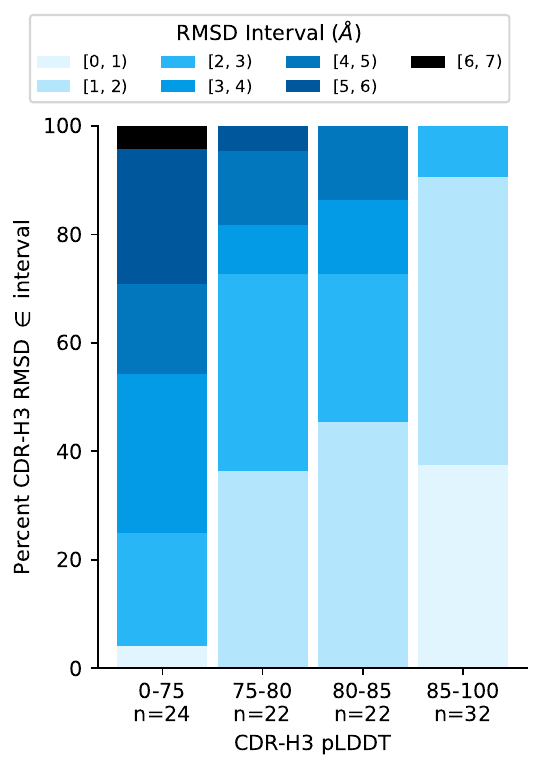}
    \caption{Left: Structure predicted by ABodyBuilder3, with colouring indicating the pLDDT uncertainty estimate. The ground truth (7T0J) is shown in grey. Right: Distribution of CDRH3 RMSD across different bins of the CDRH3 pLDDT score.}
    \label{fig:uncertainty}
\end{figure*}

We use the Structural Antibody Database (SAbDab)~\cite{dunbar2014sabdab}, a dataset of experimentally resolved antibody structures, to train our models on all available data up to January 2024.
We perform an initial filtering to remove nanobodies, structures with resolution above 3.5{\AA}, and outliers more than 3.5 standard deviations from the mean for any of the six summary statistics given by ABangle~\cite{dunbar2013abangle}.
Furthermore, we filter out ultra-long CDRH3 loops, which predominantly come from bovine antibodies \cite{de2015structural} by removing any sequence with a CDRH3 of over $30$ residues. 
We label residues using IMGT numbering generated via ANARCI \cite{dunbar2016anarci}. 
In an attempt to remove potential structural outliers, we also remove antibodies from species which occur less than $15$ times in SAbDab. 

For both the first and second stage of training we select weights based on the lowest validation loss. 
We use a validation set of $150$ structures and a test set of $100$ structures, which are significantly larger than those used in ABodyBuilder2 and lead to a more robust assessment of model capabilities. We retain the original ABodyBuilder2 test set of 34 structures as a subset of our test set to allow for direct comparisons. 
As a primary interest is the modelling of antibodies with high humanness in the context of therapeutic antibody development, we constrain the validation and additional test structures to be annotated as human.
We require a resolution below 2.5{\AA} and a CDRH3 length of less than 22 for the selection of our validation structures.
Furthermore, we remove any structures from the training data that share an identical sequence in any of the CDR regions with any of the validation or test sets.

We consider two physics-based refinement strategies, OpenMM~\cite{openmm} and YASARA~\cite{yasara}, to fix stereochemical errors and provide realistic structures. 
We find that minimization in the YASARA2 forcefield~\cite{krieger2009improving} in explicit water leads to improved accuracy of all regions, particularly in the framework.
Further details and comparisons between the minimization methods are given in Appendix~\ref{app:refinement}.

In Table~\ref{table:rmsd}, the first three rows show a comparison of the original ABodyBuilder2 model with our baseline model obtained with our improved implementation and dataset curation. 
We give the root mean squared deviation (RMSD) for each region of the variable domain, and provide results with both refinement strategies.
Note here that ABodyBuilder2 predictions are obtained by taking the closest structure to the mean of an ensemble of four models. This ensemble of models is selected from ten distinct trainings of which six models are then discarded. By comparison, our baseline consists of a single model without any need for model selection and ensemble prediction.

\section{Language model representation}
Inspired by the success of language model embeddings being used to model protein structure, e.g.~\cite{esmfold,igfold}, we experiment with replacing the one-hot-encoding used as the residue representation in ABodyBuilder2 with a language model embedding. 
We use the ProtT5 model~\cite{elnaggar2021prottrans}, an encoder-decoder text-to-text transformer model~\cite{t5} pretrained on billions of protein sequences, to generate a residue level embedding of each antibody. 
As this language model was trained on single chains, we embed the heavy and light chain separately by passing them through the ProtT5 model, and concatenate their residue representations to obtain a per-residue embedding of the full variable region.
We also explored antibody-specific models such as the paired IgT5 and IgBert models~\cite{igt5}, but ultimately found that general protein language models achieved higher performance.
This might be because antibody language models introduce potential dataset contamination and overfitting during the language model pre-training.
To train our language model-based structure prediction model, all parameters are kept identical to ABodyBuilder2, except for a lower initial learning rate of $5\cdot 10^{-4}$, and a reduction of the minimum learning rate to 0 in the scheduler, which we found to improve stability on learning rate resets.

In Table~\ref{table:rmsd}, we show the performance of our ABodyBuilder3 model, comparing the one-hot encoding with the ProtT5 embedding representation which we denote as ABodyBuilder3-LM.
One can observe a small reduction in RMSD using the language model representation, notably in the modelling of the CDRH3 and CDRL3 loops.

\begin{table*}[h!]
    \centering
    \begin{tabular}{lcccccccc}
        \toprule
        & CDRH1 & CDRH2 & CDRH3 & Fw-H & CDRL1 & CDRL2 & CDRL3 & Fw-L \\
        \midrule
        ABodyBuilder2 & 0.84 & 0.73 & 2.54 & 0.56 & 0.55 & 0.36  & 0.88 & 0.53  \\
        Baseline (OpenMM) & 0.92 & 0.75 & 2.53 & 0.60 & 0.67 & 0.35 & 0.96 & 0.58 \\
        Baseline (Yasara) & 0.90 & 0.74 & 2.49 & 0.59 & 0.58 & 0.37 & 0.92 & 0.57 \\
        ABodyBuilder3 & 0.87 & 0.70 & 2.42 & 0.58 & 0.61 & 0.39 & 0.93 & 0.58 \\
        ABodyBuilder3-LM & 0.87 & 0.75 & 2.40 & 0.57 & 0.59 & 0.37 & 0.89 & 0.58 \\
        \bottomrule
    \end{tabular}
    \caption{Modelling accuracy as measured by mean RMSD in Angstroms, given for each CDR loop and framework region. Here ABodyBuilder2 uses an ensemble of predictions and the reported accuracy is for the closest to the mean of four models, while all other models are obtained from a single prediction.}
    \label{table:rmsd}
\end{table*}

\begin{table*}[h!]
    \centering
    \begin{tabular}{lcccccccc}
        \toprule
        & CDRH1 & CDRH2 & CDRH3 & Fw-H & CDRL1 & CDRL2 & CDRL3 & Fw-L \\
        \midrule
        ABodyBuilder2 & 0.41 & 0.38 & 0.57 & 0.50 & 0.47 & 0.48 & 0.72 & 0.40   \\
        ABodyBuilder3 & 0.58 & 0.26 & 0.61 & 0.48 & 0.60 & 0.20 & 0.68 & 0.67 \\
        ABodyBuilder3-LM & 0.69 & 0.36 & 0.73 & 0.39 & 0.72 & 0.52 & 0.68  & 0.58 \\
        \bottomrule
    \end{tabular}
    \caption{Pearson correlation between average uncertainty prediction for a region and the corresponding mean RMSD. Uncertainties for ABodyBuilder2 are derived from an ensemble of four models, while all ABodyBuilder3 uncertainties are directly predicted by a pLDDT head.}
    \label{table:pearson}
\end{table*}

\section{Uncertainty estimation}
The ABodyBuilder2 model uses an ensemble of four models to provide a confidence score from the diversity between predictions. 
This approach has an increased computational burden, as multiple models are required both at training and inference time. 
We instead estimate the intrinsic model accuracy by predicting the per-residue lDDT-C$\alpha$ scores~\cite{lddt}, as implemented in the AlphaFold2 model~\cite{jumper2021highly}.
This introduces a very small increase in the number of parameters, but circumvents the need for an ensemble of models.
The pLDDT is obtained from the final single representation of the structure module and predicts a projection into 50 bins by a multilayer perceptron with softmax activation. Training is achieved by discretising the predicted structure with per-residue lDDT-C$\alpha$ against the ground truth structure and computing the cross-entropy loss, which is added to the original ABodyBuilder2 loss with a weight of 0.01. 
A pLDDT score for the full variable domain, or for specific regions, is obtained as an average of the corresponding per-residue pLDDT scores.

In Table~\ref{table:pearson}, we give the Pearson correlation between the pLDDT score and the RMSD, averaged over each region of the variable domain.
The ABodyBuilder2 model, with an uncertainty score obtained from the ensemble model, has lower correlation with RMSD than our pLDDT score.
It is interesting to note here that the ABodyBuilder3-LM model, which uses ProtT5 embeddings as input, achieves a higher correlation than the one-hot encoding representation model, notably in the modelling of the CDRH3 uncertainty.
We note however that when considering the Spearman correlation, shown in Appendix~\ref{app:refinement}, the difference between models is less marked.
We provide a guideline for thresholding pLDDT for modelling the CDRH3 region in Figure~\ref{fig:uncertainty} (right), applied here on the full test set. 
Incorporating a threshold of a pLDDT above $85$, we retain approximately $32\%$ of structures, with over $80\%$ of those retained having a CDR-H3 RMSD below 2{\AA}.

\section{Conclusions}
In this article, we present ABodyBuilder3, a state-of-the-art antibody structure prediction model based on ABodyBuilder2.
We incorporated several improvements to the implementation, notably enhancing hardware acceleration through vectorisation, which significantly improve the scalability of our model. 
We also made changes to the data processing and structure refinement that lead to more accurate predictions.

In addition, we show how leveraging a language model representation of the antibody sequence can improve the modelling of CDRH3.
Finally, we demonstrate how the use of pLDDT head, combined with protein language model embeddings, can be used as a substitute for an ensemble of models approach, which require substantially more training and inference compute.

It would be interesting to explore the use of self-distillation, which has shown to improve accuracy in general protein structure prediction models~\cite{af2}, by pre-training our model on a large dataset of synthetic structures predicted from the paired Observed Antibody Space~\cite{OAS1,abb2data}.
To even further improve the accuracy of the predictions and of the uncertainty estimates, one could also consider combining the pLDDT approach introduced in this article with an ensemble of models, though this would be at the cost of increased training and inference compute.

We release the code\footnote{\href{https://github.com/Exscientia/ABodyBuilder3}{github.com/Exscientia/ABodyBuilder3}} and model weights for ABodyBuilder3~\cite{zenodo}.

\section*{Acknowledgements}
We are grateful to Brennan Abanades for numerous helpful discussions and advice throughout this project, as well to Constantin Schneider, Claire Marks and Aleksandr Kovaltsuk for useful comments.

\bibliographystyle{icml2023}
\bibliography{bibliography}

\newpage
\appendix
\onecolumn

\section{Structure refinement}
\label{app:refinement}
\begin{figure*}[ht]
    \centering
    \includegraphics[width=1.0\linewidth]{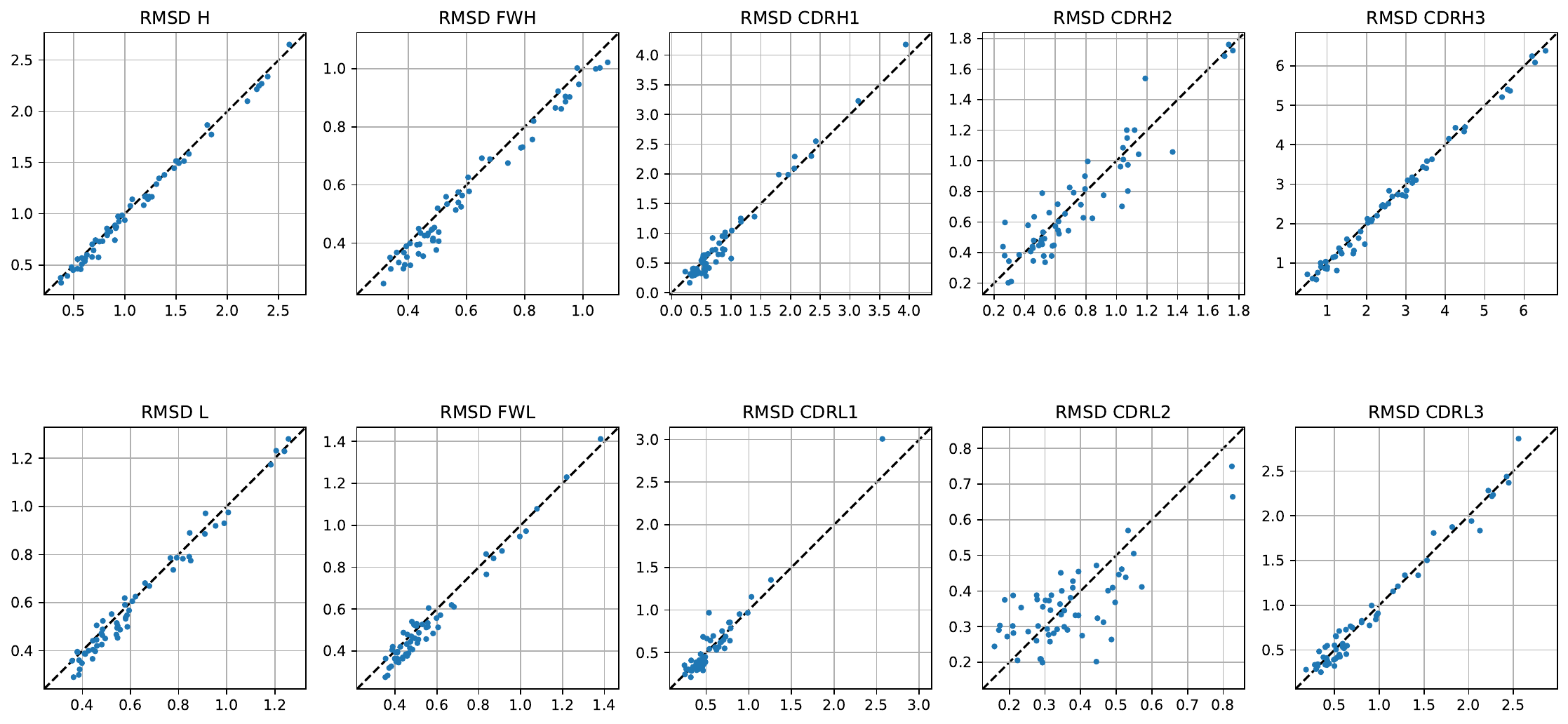}
    \caption{YASARA2 refinement (x-axis) compared to OpenMM refinement (y-axis).}
    \label{fig:yasara}
\end{figure*}

Most approaches to protein or antibody structure prediction, including AlphaFold2~\cite{af2}, AlphaFold2-Multimer~\cite{afm}, and ABodyBuilder2~\cite{abb2}, that predominantly rely on deep learning also use a final physics-based refinement step to fix stereochemical errors and provide realistic structures. In most cases, this refinement takes the form of an in-vacuo minimization that may neglect important aspects of the real system in favor of expediency. To test the influence of different types of minimization on the quality of ABodyBuilder3 output structures, we compare an OpenMM refinement with a refinement using YASARA2, shown in Figure~\ref{fig:yasara}. These results confirm that while minimization in vacuo is sufficient to improve many structures, a minimization in the YASARA2 forcefield~\cite{krieger2009improving, yasara} in explicit water allows for further improvements across all the regions, particularly in the framework, whilst improving model quality according to the z-score produced by YASARA2.

\section{Spearman correlation of uncertainties}
\label{app:spearman}
We show the Spearman correlation between uncertainty predictions and RMSD in Table~\ref{table:spearman}.

\begin{table*}[h!]
    \centering
    \begin{tabular}{lcccccccc}
        \toprule
        & CDRH1 & CDRH2 & CDRH3 & Fw-H & CDRL1 & CDRL2 & CDRL3 & Fw-L \\
        \midrule
        ABodyBuilder2 & 0.45 & 0.30 & 0.75 & 0.50 & 0.56 & 0.21 & 0.73 & 0.42  \\
        ABodyBuilder3 & 0.48 & 0.23 & 0.63 & 0.50 & 0.37 & 0.01 & 0.59 & 0.59 \\
        ABodyBuilder3-LM & 0.48 & 0.40 & 0.73 & 0.27 & 0.43 & 0.20 & 0.60 & 0.53 \\
        \bottomrule
    \end{tabular}
    \caption{Spearman correlation between average uncertainty prediction for a region and the corresponding mean RMSD.}
    \label{table:spearman}
\end{table*}

\end{document}